\nonstopmode
\documentclass[twocolumn]{aastex6} 

\makeatletter\renewcommand\theenumi{\@alph\c@enumi}\makeatother

\def\hii{\relax \ifmmode {\mbox H\,\textsc{ii}}\else H\,{\scshape ii}\fi}

\begin{document}


\title{Evidence of ongoing radial migration in NGC~6754: \\Azimuthal variations of the gas properties}
\shorttitle{Azimuthal variations of the gas properties in NGC~6754} 

\shortauthors{L.~S\'anchez-Menguiano et al.}

\author{L.~S\'anchez-Menguiano\altaffilmark{1,2} \and S.~F.~S\'anchez\altaffilmark{3} \and D.~Kawata\altaffilmark{4} \and L.~Chemin\altaffilmark{5} \and I.~P\'erez\altaffilmark{2} \and T.~Ruiz-Lara\altaffilmark{2} \and P.~S\'anchez-Bl\'azquez\altaffilmark{6} \and L.~Galbany\altaffilmark{7,8} \and J.~P.~Anderson\altaffilmark{9} \and R.~J.~J.~Grand\altaffilmark{10,11} \and I.~Minchev\altaffilmark{12} \and F.~A.~G\'omez\altaffilmark{13}}
\email{lsanchez@iaa.es}

\altaffiltext{1}{Instituto de Astrof\'isica de Andaluc\'ia (CSIC), Glorieta de la Astronom\'ia s/n, Aptdo. 3004, E-18080 Granada, Spain}
\altaffiltext{2}{Dpto. de F\'isica Te\'orica y del Cosmos, Universidad de Granada, Facultad de Ciencias (Edificio Mecenas), E-18071 Granada, Spain}
\altaffiltext{3}{Instituto de Astronom\'ia, Universidad Nacional Aut\'onoma de M\'exico, A.P. 70-264, 04510, M\'exico, D.F.}
\altaffiltext{4}{Mullard Space Science Laboratory, University College London, Holmbury St. Mary, Dorking, Surrey, RH5 6NT, UK}
\altaffiltext{5}{INPE/MCT, Divis$\tilde{a}$o de Astrof\'isica, S. J. dos Campos, Brazil}
\altaffiltext{6}{Instituto de Astrof\'isica, Pontificia Universidad Cat\'olica de Chile, Av. Vicu\~na Mackenna 4860, 782-0436 Macul, Santiago, Chile}
\altaffiltext{7}{Millennium Institute of Astrophysics, Chile}
\altaffiltext{8}{Departamento de Astronom\'ia, Universidad de Chile, Camino El Observatorio 1515, Las Condes, Santiago, Chile}
\altaffiltext{9}{European Southern Observatory, Alonso de C\'ordova 3107, Casilla 19001, Santiago, Chile}
\altaffiltext{10}{Heidelberger Institut f\"ur Theoretische Studien, Schloss-Wolfsbrunnenweg 35, 69118 Heidelberg, Germany}
\altaffiltext{11}{Zentrum f\"ur Astronomie der Universit\"at Heidelberg, Astronomisches Recheninstitut, M\"onchhofstr. 12-14, 69120 Heidelberg, Germany}
\altaffiltext{12}{Leibniz-Institut f\"{ur} Astrophysik Potsdam (AIP), An der Sternwarte 16, D-14482, Potsdam, Germany}
\altaffiltext{13}{Max-Planck-Institut f\"ur Astrophysik, Karl-Schwarzschild-Str. 1, D-85748, Garching, Germany}





\begin{abstract}
Understanding the nature of spiral structure in disk galaxies is one of the main, and still unsolved questions in galactic astronomy. However, theoretical works are proposing new testable predictions whose detection is becoming feasible with recent development in instrumentation. In particular, streaming motions along spiral arms are expected to induce azimuthal variations in the chemical composition of a galaxy at a given galactic radius. In this letter we analyse the gas content in NGC~6754 with VLT/MUSE data to characterise its 2D chemical composition and H$\alpha$ line-of-sight velocity distribution. We find that the trailing (leading) edge of the NGC~6754 spiral arms show signatures of tangentially-slower, radially-outward (tangentially-faster, radially-inward) streaming motions of metal-rich (poor) gas over a large range of radii. These results show direct evidence of gas radial migration for the first time. We compare our results with the gas behaviour in a $N$-body disk simulation showing spiral morphological features rotating with a similar speed as the gas at every radius, in good agreement with the observed trend. This indicates that the spiral arm features in NGC~6754 may be transient and rotate similarly as the gas does at a large range of radii.
\end{abstract}


\section{Introduction}\label{sec:intro}

Spiral galaxies are dynamical systems in which their main components, gas and stars, undergo outward and inward radial excursions. The reason for this so-called radial migration has been broadly investigated. For instance, \citet{sellwoodbinney2002} proposed that stars and gas close to the co-rotation resonance associated with the spiral structure experience large changes in their radial positions. However, this mechanism depends on the nature of the spiral structure. In classic density wave theory \citep{linshu1964}, the spiral arms are assumed to be rigidly-rotating, long-lived patterns that, as a consequence, present a unique co-rotation radius at which the spiral arms and stars rotate at the same speed (and at which radial migration can occur). \citet{juliantoomre1966} proposed a different theory based on transient and recurrent patterns that form through local instabilities which are swing amplified into spiral arms. \citet{sellwoodbinney2002} claimed that only transient spiral arms can induce long lasting radial migration, otherwise migrated stars are returned to their initial locations on horseshoe orbits.

Although the causes of radial migration remain still unclear, several simulations have been attempted to understand this important phenomenon in galaxy evolution, finding that both gas and stars might be affected significantly by it \citep[e.g.][]{minchev2014, grand2015a}. In particular, recent simulations have shown that radial migration can induce azimuthal variations of the stellar metallicity distribution \citep{dimatteo2013,grand2016}. Regarding the gas content, streaming motions along the spiral arms have been found \citep[][]{grand2015a, baba2016}, which could also produce azimuthal variations of the gas abundance.

Despite significant research attempting to understand these migration processes, there is currently little observational evidence for this mechanism in galaxies \citep{magrini2016}. In particular, only a few studies have analysed possible azimuthal variations in the gas abundance distribution of spiral galaxies \citep[e.g.][]{kennicutt1996, martin1996, cedrescepa2002, rosalesortega2011, cedres2012, li2013}, without finding variations related to the presence of the arms. The lack or presence of this kind of azimuthal trends would not only allow us to assess the importance of radial migration driven by spiral arms in shaping these galaxies, but also to shed light onto the nature of the spiral structure itself. The advent of a new generation of high spatial resolution instruments such as VLT/MUSE or ALMA can provide the data quality needed to bring such constraints within reach.

In this letter we use VLT/MUSE data of NGC~6754, a galaxy that has shown hints of radial migratory processes \citep{sanchez2015a}. NGC~6754 is an isolated barred Sb galaxy slightly inclined ($i=62.6\degr$, ${\rm PA=86\degr}$, see Sect.~\ref{sec:analysis}) located at a redshift of 0.0109 \citep[47 Mpc assuming a WMAP9 cosmology,][]{hinshaw2013}. Here, we undertake a deeper study using 2-D information to analyse the residual gas abundance (after removing the azimuthally averaged radial abundance) and velocity maps of NGC~6754 to search for possible evidence of gas radial migration as proposed in simulations.

The structure of the letter is organized as follows. Sections~\ref{sec:data} and \ref{sec:analysis} provide a description of the data and the analysis required to derive the oxygen abundance and line-of-sight (LOS) velocity residual distributions. The presentation of the results and a comparison with simulations focused on the gas content is given in Section~\ref{sec:results}. Finally, Section~\ref{sec:conclusions} outlines the main conclusions.


\section{Observations and data reduction}\label{sec:data}

Observations of NGC~6754 were carried out using the MUSE instrument \citep{bacon2010} as part of the AMUSING Survey \citep{galbany2016}. MUSE is an integral-field spectrograph that provides a large field-of-view (FoV) of $1'\times1'$ with a sampling of $0.2''$/spaxel. The covered wavelength range spans between ${\rm4750-9300\,\AA}$, with a spectral sampling of ${\rm1.25\,\AA}$ and a spectral resolution between $1800-3600$. 

Observations were split into two different pointings consisting of three exposures of 900 seconds and covering the eastern and western parts of the galaxy under seeing conditions of $0.8''$ ($180$~pc) and $1.8''$ ($410$~pc), respectively. The mosaic was corrected by the effects of Galactic extinction. Information about the data reduction can be found in \citet{galbany2016}. The final dataset comprises almost 200k individual spectra with a FoV of $\sim2'\times1'$, covering the entire galaxy up to two effective radius ($\rm r_e\sim6.2\;kpc$, S\'anchez-Menguiano et al. in prep).


\section{Analysis}\label{sec:analysis}

In this study we analyse the full 2-D information of the ionised gas provided by the data avoiding any binning schemes \citep[following][]{sanchezmenguiano2016}. In this section we briefly summarize the procedure followed to select the spaxels, analyse their individual spectra and derive the corresponding LOS velocity and oxygen abundance distributions and their residual maps. Both distributions were later deprojected using the position and inclination angles quoted before, derived by fitting ellipses of variable ellipticity and PA to the outermost isophotes of the galaxy in a $g$-band image recovered from the data.

\begin{figure*}
\begin{center}
\resizebox{\hsize}{!}{\includegraphics{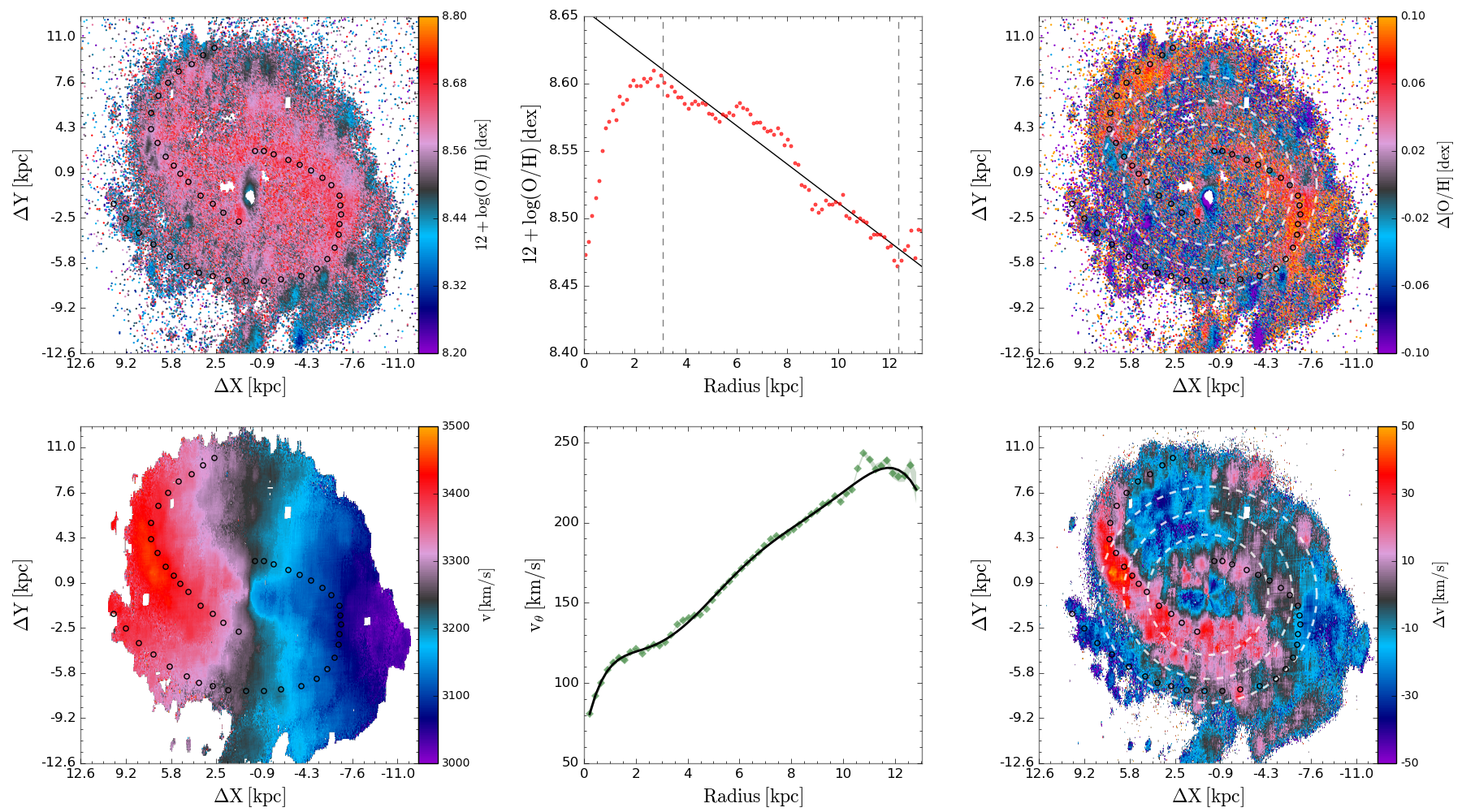}}
\caption{2-D deprojected distributions of the gas oxygen abundance (top left), residual abundance (top right), H$\alpha$ LOS velocity (bottom left) and LOS velocity residuals (bottom right). The position $\rm\Delta X=\Delta Y=0$ corresponds to the center of the galaxy. The location of the spiral arms is marked with open circles. Dashed white circles indicate the three radial positions ${\rm R=4.5,\,6.3\;and\;8.1\:kpc}$ of the azimuthal profiles shown in Fig.~\ref{fig:plot_azi}. Top middle panel shows the metallicity gradient (black) fitted to the azimuthally averaged radial abundances (red) in the linear regime (delimited by the dashed vertical lines, see \citealt{sanchezmenguiano2016} for further information). Bottom middle panel shows the ionized gas rotation curve (green) with the $1\sigma$ formal errors (light green), together with its smoothed version (polynomial fit, solid line).}
\label{fig:plot_obs}
\end{center}
\end{figure*}

\subsection{Derivation of the residual abundance map}\label{sec:analysis1}

To measure the emission line fluxes needed to derive oxygen abundances, we first remove the stellar population contribution. We model both the continuum emission and emission lines using Pipe3D, as described in \citet{sanchez2016a}. Briefly, Pipe3D fits each spectrum by a linear combination of single stellar population templates after correcting for the appropriate systemic velocity and velocity dispersion and taking into account the effects of dust attenuation \citep{cardelli1989}. Once the stellar component is subtracted, then Pipe3D measures the emission lines performing a multi-component fitting using a single Gaussian function per emission line plus a low order polynomial function. 

We select the spaxels associated with star formation using well-known diagnostic diagrams, in particular, that proposed by \citet[][]{baldwin1981} based on the \mbox{[\ion{N}{2}]~$\lambda6584$/H$\alpha$} and \mbox{[\ion{O}{3}]~$\lambda5007$/H$\beta$} line ratios, together with the \citet{kewley2001} demarcation line. We have also made use of the so-called WHAN diagram \citep[W$_{\rm H\alpha}$ versus \mbox{[\ion{N}{2}]/H$\alpha$},][]{cidfernandes2011}, based on the equivalent width (EW) of H$\alpha$. However, we have been more restrictive in the EW range, using a limit of ${\rm6\;\AA}$ to guarantee a higher S/N of the emission lines and remove any contribution coming from the diffuse nebular emission. 

To obtain the oxygen abundance distribution we adopt the empirical calibrator based on the O3N2 index described in \citet{pettini2004} and the calibration proposed by \citet{marino2013}. We must note that these empirical calibrators are based on spectroscopic data of integrated \hii\, regions. The seeing conditions of the observations ($0.8''/1.8''$) do not allow us to spatially resolve the \hii\, regions at the redshift of the galaxy, making possible the use of this calibrator for the analysis.

Finally, the residual map is derived by subtracting the azimuthally averaged radial abundance to the observed distribution. These averaged values are measured as the median of the abundances in the area of the disk out of the spiral arms (assuming an arm width of $6''$ to ensure that no arm contamination is considered in the average). 

\subsection{Derivation of the residual velocity map}\label{sec:analysis2}

Pipe3D also provides LOS velocities from which we can derive gas velocity maps of the galaxy for each analysed line. We focus the analysis on the measurements of the H$\alpha$ emission line, as it represents the strongest detected line.

The rotation curve, $v_\theta$, and the radial velocity component, $v_R$, were obtained by least-square fits of $v_{\rm sys}+(v_\theta\cos\theta+v_R\sin\theta)\sin\,i$ to the observed velocity field, $v_{\rm obs}$, assuming constant coordinates of the mass centre, systemic velocity ($v_{\rm sys}$), disk position angle and inclination (matching the photometric parameters). The best-fit systemic velocity corresponding to these parameters is $3247\pm9$~km/s. We sampled the galactocentric radius using 1\arcsec-rings (230~pc-rings), roughly similar to the seeing of the observations. With that sampling, rings are uncorrelated, and the least-square fits were performed with at least 70 degrees-of-freedom, so that $v_R$ and $v_\theta$ are very well constrained (average $1\sigma$ formal error of 1.3~km/s). The axisymmetric model velocity field, $v_{\rm mod}$, is then the projection of a smoothed version of v$_\theta$ only and does not contain the contribution from the fitted $v_R$. Therefore, the residual LOS velocities defined as $\Delta v=v_{\rm obs}-v_{\rm mod}$, trace the observed non-circular radial motions, as well as the local departures from axisymmetry of both $v_R$ and $v_\theta$ (streaming motions caused by the bar, spiral perturbations, etc.).


\section{Results and discussion}\label{sec:results}

We have derived the residual maps of the oxygen abundance and the H$\alpha$ LOS velocity distributions of NGC~6754 in order to study possible asymmetries in these distributions linked to the spiral structure.

Now, we analyse these observational results, and provide a comparison to simulations in order to further interpret our findings.

\subsection{Asymmetries in the residual abundance and velocity distributions}

In Figure~\ref{fig:plot_obs} we present the results of our observational analysis. The top left-hand panel displays the 2-D distribution of the oxygen abundance and top right-hand panel displays the residuals after subtracting the azimuthally averaged radial abundances (top middle). In this panel we can see clear differences associated with the spiral arms (black markers). The bottom left-hand panel displays the 2-D distribution of the LOS gas velocity and bottom right-hand panel displays the residuals after subtracting the LOS projection of the derived rotation model (bottom middle). This map shows a feature associated to the eastern (left-hand side in this panel) spiral arm, with its leading part presenting higher velocity residuals than the trailing side. The reason why this feature is not observed (so clearly) for the western arm may lie in the lower spatial resolution of the data covering this half of the galaxy due to the worse seeing conditions (see Sect.~\ref{sec:data}). Therefore, below we will focus on the eastern spiral arm.

\begin{figure}
\begin{center}
\resizebox{\hsize}{!}{\includegraphics{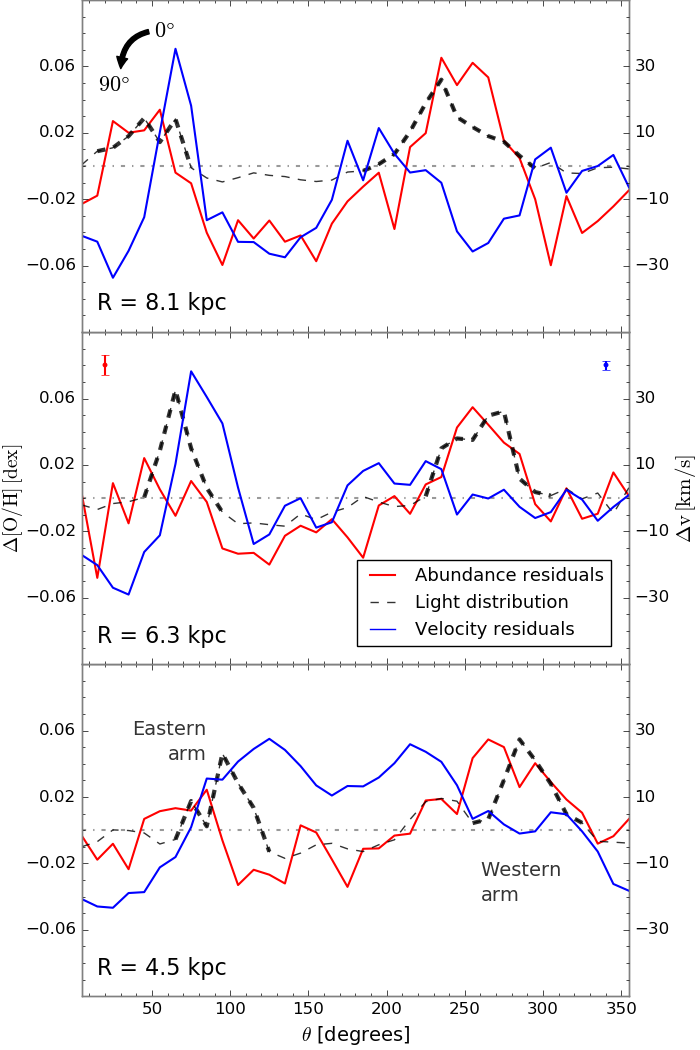}}
\caption{Azimuthal profiles of the light (black dashed), oxygen abundance residuals (blue, left-hand y-axis) and H$\alpha$ LOS velocity residuals (red, right-hand y-axis) at three different radii (shown in Fig.~\ref{fig:plot_obs}). The position of the spiral arms is marked with the bold dashed line. Mean errors in the azimuthal profiles are denoted by the vertical lines in the middle panel. The angles are measured counter-clockwise from the positive Y-axis in Figure~\ref{fig:plot_obs}.}
\label{fig:plot_azi}
\end{center}
\end{figure}

Figure~\ref{fig:plot_azi} displays the azimuthal profiles of both residuals for three different $2\arcsec$-wide annuli (centered at \mbox{$\rm R=4.5/6.3/8.1\:kpc$}, i.e. $\rm 0.7/1.0/1.3\:r_e$) with a sampling of 10\degr. Around the eastern arm, the azimuthal residual LOS velocity profiles show a peak located just after the peak in the light distribution (leading side of the arm), with an amplitude between $\sim28-38$~km/s, and a minimum just before the light peak (trailing side). Remarkably, these maxima (minima) in the velocity profile appear together with a decrement (increment) in the azimuthal residual abundance profile at all the radii, with a total amplitude (peak-to-peak) up to 0.09 dex.

\begin{figure*}
\begin{center}
\resizebox{0.8\hsize}{!}{\includegraphics{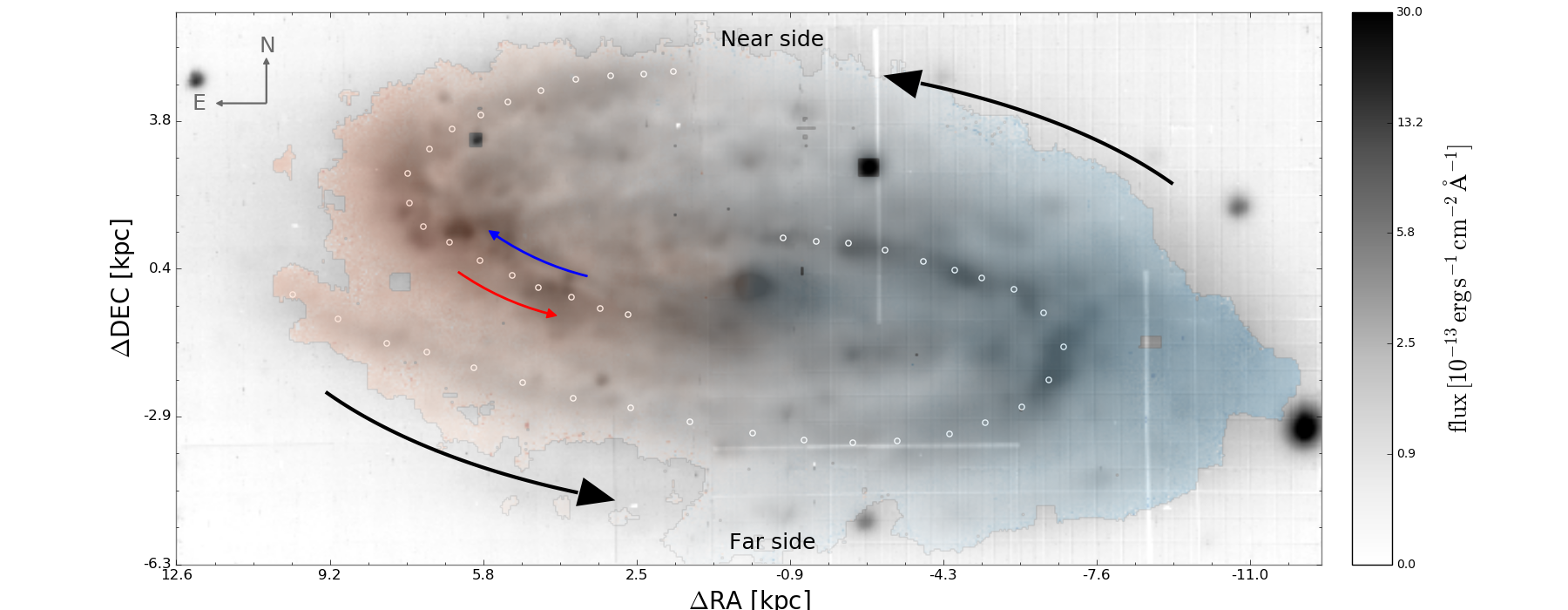}}
\caption{$g$-band image of NGC~6754 superimposed to the observed H$\alpha$ velocity map, with red (blue) color denoting the receding (approaching) part of the galaxy. The black arrows indicate the direction of rotation (assuming trailing spiral arms). Radially-inward/tangentially-faster and radially-outward/tangentially-slower motions of the gas along the arms (outlined by open circles) are indicated with red and blue arrows, respectively (see text for a detailed explanation).}
\label{fig:plot_orientation}
\end{center}
\end{figure*}

In order to interpret these results, in Fig.~\ref{fig:plot_orientation} we show a $g$-band image of NGC~6754 superimposed to its observed H$\alpha$ velocity map. We consider that the North part of the galaxy (upper-side in the figure) is closer to us under the assumption of trailing spiral arms \citep[][]{hubble1943} and taking into account that the eastern part is receding from us. In that case, positive (negative) residual velocities indicate radially-inward (outward) motions of the gas and tangentially-faster (slower) motions of the gas for the eastern spiral arm. Thus, the positive velocity residuals displayed by NGC~6754 in a wide extension of the leading part of the eastern arm can be interpreted either as gas moving radially-inward, gas moving tangentially-faster, or a combination of both. Following a similar reasoning, the negative velocity residuals in the trailing part can be the result of gas moving radially-outward, gas moving tangentially-slower, or both. The asymmetries found in the metallicity residuals are in agreement with a transport of metal-rich gas from the inner disk towards the outer regions at the trailing-side of the spiral arm and more metal-poor gas from the outer disk towards the inner ones at the leading-side, which is strikingly consistent with the velocity asymmetries mentioned above. These trends are observed at all three radii, which indicates the strong evidence of the radial migration happening in a large radial range.

\subsection{Comparison with simulations}

In order to investigate the behaviour of the gas from a theoretical perspective, in this section we study an isolated Milky Way-sized disk galaxy simulated using the $N$-body smoothed particle hydrodynamics (SPH) GCD+ code. Details of the code are available in \citet[][]{kawatagibson2003} and \citet[][]{kawata2013}. In particular, we analyse the simulation labelled K14 in \citet{grand2015b}, which is similar in size to NGC~6754. For this analysis, we focus on a single snapshot of the galaxy that displays a clear bar and two-armed spiral structure, and shows clear radial migration of the gas around the spiral arm \citep{kawata2014,grand2015a}. The rotation axis and the angle of the spiral arm have been chosen to match the characteristics of NGC~6754.

\begin{figure*}
\begin{center}
\resizebox{0.8\hsize}{!}{\includegraphics{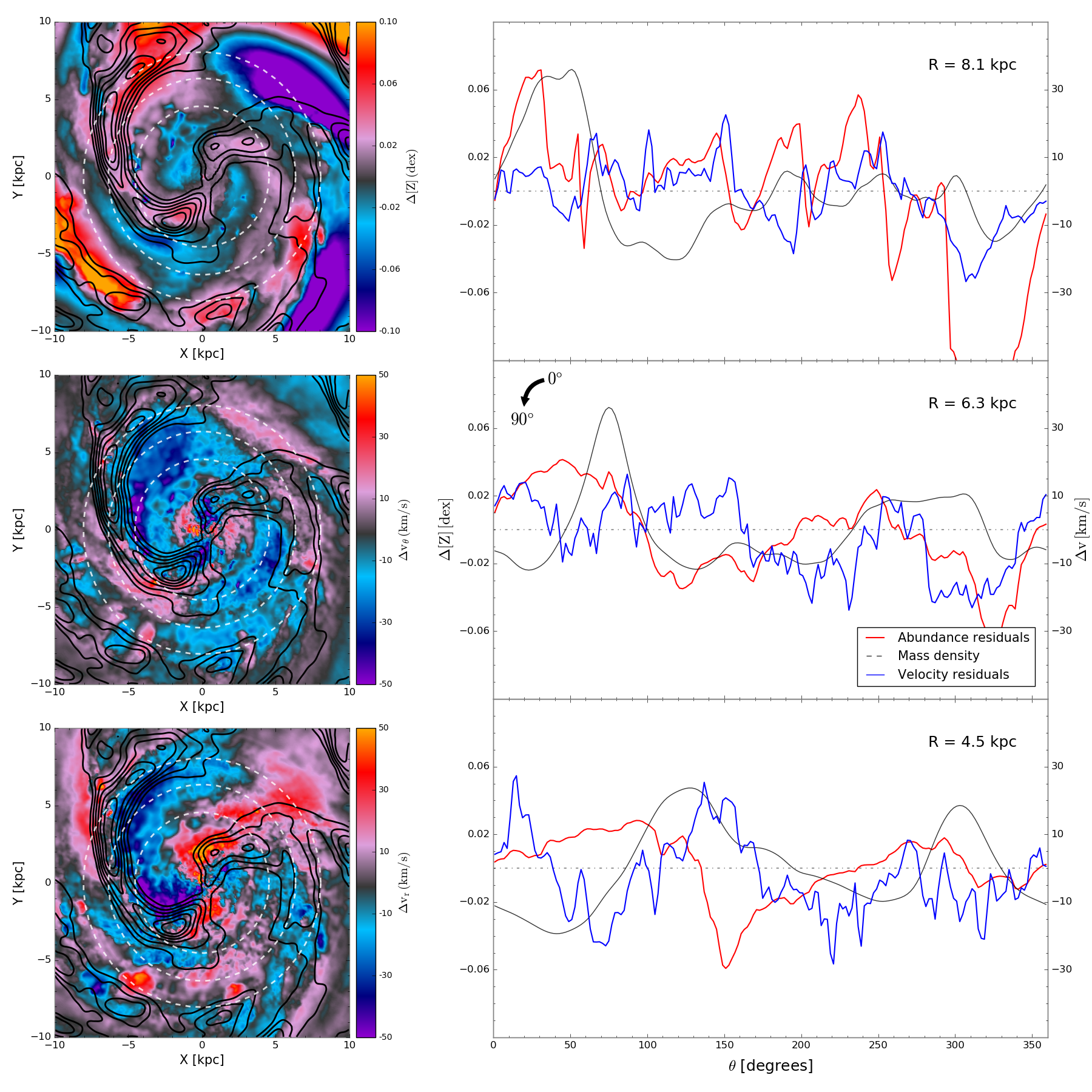}}
\caption{Chemodynamical results from simulations. {\it Left-hand panels:} Face-on maps of the residual gas metallicity (top), tangential residual velocity field (middle) and radial velocity field (bottom) of the simulated galaxy. Contour maps represent the mass density and the dashed white circles indicate the radial position of the azimuthal profiles. {\it Right-hand panel:} Same plots as in Fig.~\ref{fig:plot_azi}, but here produced from our simulated galaxy.}
\label{fig:plot_simu}
\end{center}
\end{figure*}

The azimuthal analysis of the gas content in this simulated galaxy was performed in a similar way as in NGC~6754. Figure~\ref{fig:plot_simu} (right-hand panel) shows the azimuthal profiles of the mass density (black), residual metallicity (red) and residual LOS velocity (blue) for assumed $\rm i=63\degr$. The mass density distribution shows a clear peak below $200\degr$, which represents the eastern spiral arm. As the western (right-hand side) arm is much weaker, especially in the outer radii in this particular simulation snapshot, we will focus on the stronger eastern (left-hand side) arm to compare with the trends observed in NGC~6754. Consistent with the observations, the residual LOS velocity (metallicity) is lower (higher) on the trailing side of the spiral, while the trend is reversed on the leading side, i.e., residual LOS velocity (metallicity) is higher (lower) than the average values at the three analysed radii matching those chosen for Figure~\ref{fig:plot_azi}. 

The simulations allow us to analyse separately the radial and tangential components of the LOS velocity. The left-hand panels of Fig.~\ref{fig:plot_simu} display face-on maps of the residuals of the gas metallicity (top) and tangential velocity field (middle) from the analysed simulation after subtracting the azimuthally averaged values at each radius. In addition, the radial velocity field map is shown in the bottom panel. We can see that the tangential velocity is slower on the trailing side of the eastern spiral arm and faster on the leading edge with respect to the general rotation, whereas the radial velocity points outward on the trailing side and inward on the leading edge. The metallicity map shows the presence of metal-rich (poor) gas particles on the trailing (leading) side of the spiral arm. The combination of both tangential and radial velocity behaviours creates a streaming motion that, along with the well-known negative gas metallicity profile of disks \citep[e.g.][]{searle1971, martin1992, sanchez2014}, cause metal-rich (poor) gas to move towards the outer (inner) regions at the trailing (leading) side of the spiral arm in a large radial range.

\section{Conclusions}\label{sec:conclusions}

In this letter we have analysed azimuthal variations of the gas metallicity and velocity residuals in NGC~6754 using AMUSING data (VLT/MUSE). 

Studying the eastern spiral arm, located in the half of the disk with data of higher spatial resolution, the residual LOS velocity distribution shows a maximum at the leading side of the spiral arm (larger receding component in the velocity), with an amplitude of $\sim30$~km/s, and a minimum at the trailing edge. The residual abundance distribution is positive at the trailing side and negative at the leading edge, with a total amplitude (peak-to-peak) of $\sim0.1$~dex. We note that although NGC~6754 seems to have a companion close by (2MASXJ19112166-5037339), the linear distance between them (26 Mpc) suggests that our results cannot be a consequence of interaction \citep[see also][]{demello1996}.

The spatial resolution of the data is crucial in the analysis of azimuthal variations. In this study we find that, in order to be sensitive to the signatures left by radial migration, we need a spatial resolution of $\sim200$~pc. Otherwise the signatures are blurred (western arm).

We have also analysed the gas content of a simulated galaxy ($N$-body+SPH) which shows a clear radial gas migration around a spiral morphological feature rotating at a similar speed as the gas at every radius, showing the same trends as our observations. In light of these results, we claim that NGC~6754 shows clear signs of ongoing radial gas migration that produces motions of metal-rich gas towards the outer regions on the trailing side of the spiral arm and metal-poor gas towards inner regions on the leading side. 

The analysed simulations show amplitudes for the residual LOS velocity of $\sim10-20$~km/s, lower than those observed in NGC~6754. The derived values of the differences in the LOS velocities and metallicities in simulations depend on several factors, such as the amplitude and pitch angle of the spiral arms, the underlying metallicity gradient, amount of metal mixing, feedback, etc. Therefore, MUSE observations of various types of spiral galaxies can provide further constraints on the simulation sub-grid physics.

The fact that these streaming motions are observed in a wide radial range of the spiral arm (at least 4~kpc) puts strong constraints on the nature of the spiral structure. Azimuthal variations in the velocity distribution across a classical density wave-like spiral arm have been suggested by some authors \citep{minchevquillen2008,chemin2016,pasetto2016}. However, the gas motion should show a clear offset from the spiral arm density peak, and the offset should strongly depend on the radius, as demonstrated in \citet{baba2016}. Also, to our knowledge, they do not find azimuthal metallicity variations. On the other hand, N-body simulations of disk galaxies commonly show spiral arm features in the morphology whose pattern speeds decrease with radius \citep[e.g.][]{wada2011, grand2012, baba2013}. These spiral features can result from the overlap of multiple modes \citep{comparetta2012}, as often claimed in simulations \citep{masset1997,quillen2011,minchev2012,sellwoodcarlberg2014}, which also induce non-linear growth of their amplitude \citep{kumamotonoguchi2016}. In this work, we have compared our results with simulations showing these spiral arm morphological features that are transient and rotate at a similar speed as the gas at every radius. Although it is not guaranteed that the nature of the spiral arms in the observed galaxy is the same as in the simulations, the consistency found supports this scenario for spiral structure formation of NGC~6754. 

In this work we present, for the first time, clear signatures of ongoing gas radial migration in which metal-poor gas clouds in the leading side of the arm are moving radially-inwards and tangentially-faster while radially-outward and tangentially-slower in the trailing edge. This is consistent with spiral morphological features whose pattern speeds decrease with radius.

Our study demonstrates the power of the MUSE data to aid our understanding of the nature of the spiral arms and radial migration. Future, high quality MUSE observations for various types of spiral galaxies will provide further constraints on the theory of the spiral arms.

\vspace{1cm}
\acknowledgements


Based on observations made with ESO Telescopes at the Paranal Observatory (programme 60.A-9329(A)). We acknowledge financial support from the Spanish {\em Ministerio de Econom\'ia y Competitividad} (AYA2012-31935), from `Junta de Andaluc\'ia' (FQM-108), and from the ConaCyt programs 180125 and DGAPA-IA100815.\\



\bibliographystyle{aasjournal} 
\bibliography{bibliography}

\end{document}